\documentclass[12pt]{article}
\title{Probabilistic foundations of quantum mechanics and quantum information}

\author{Andrei Khrennikov\\
International Center for Mathematical\\
Modeling in Physics and Cognitive Sciences,\\
MSI, University of V\"axj\"o, S-35195, Sweden\\
Email:Andrei.Khrennikov@msi.vxu.se
}

\begin{document} 
  \maketitle

\begin{abstract}
We discuss foundation of quantum mechanics (interpretations, superposition,  principle of complementarity,
locality, hidden variables) and quantum information theory. 
\end{abstract}

keywords: Foundations, contextual probability, quantum information, quantum parallelism, nonlocality

\section{INTRODUCTION}
In this paper there will be discussed foundations of quantum mechanics in connection to quantum 
information theory. Such a discussion is very important, since intensive development of models 
of quantum computing, cryptography and teleportation as well as general quantum information 
theory induced large interest to the very foundations of quantum theory. Many questions that have 
been of only theoretical (or even only philosophic) value for many years now play a fundamental role 
in applied and even technological investigations. For example, security of quantum cryptographic schemes and,
consequently, prices of corresponding market products and technologies depend
crucially on such general quantum problems as {\it complementary and nonlocality.}

We can speak about renaissance in the study of foundations of quantum theory.

And this renaissance in quantum foundations was prepared by quantum information theory, see e.g. [1], 
where more than 50 talks were presented at section "Shannon meets Bohr".

In the plenary talk of K.A. Valeev at the present conference there was given a list of fundamental
features of quantum mechanics (QM) that are very important for quantum computing (as well as cryptography,
and teleportation).

I would like to present a similar list of distinguishing features of QM
and discuss roles that those features play in our understanding of quantum reality and, in particular, 
{\it quantum information theory.} Formally K.A. Valeev and I discuss the same class of problems. However, our
viewpoints to these problems are quite different. K.A. Valeev presented the conventional viewpoint to
foundations of QM. From the conventional point of view all fundamental problems of QM (besides the problem 
of measurement) are already solved. For quantum computing (as well as cryptography and teleportation) this 
means that from the theoretical point of view this theory is well established. There are only experimental 
and technological problems. I do not think so!

In this paper, see also [2], I shall present nonconventional views to foundations of QM
and consider some consequences for quantum information theory. I start with the list of distinguishing features of QM:

\medskip

(S) This is a statistical theory of measurement.

(D) Discreteness of some fundamental physical variables.

(C) Principle of complementarity.

(NL) Nonlocality.

\section{CONVENTIONAL INTERPRETATION}

We start with (S). I think that everybody would agree that QM is a formalism that describes statistical properties of
large ensembles of very special physical systems, so called quantum systems. QM could not say anything definite about behaviour of individual quantum systems. In quantum formalism there is nothing about trajectories of individual quantum systems. We still do not know anything about dynamics of individual electrons in atom and so on. N. Bohr rightly pointed out that quantum formalism is not about quantum physical systems, but about our measurements on such systems. We can formulate this principle as {\bf Principle of contextuality.}
As a consequence of this principle, we have to take into account the whole physical arrangement of an experiment.

We continue to discuss (S). As a consequence of (S), it seems to be natural to assume that
a quantum state should describe statistical properties of an ensemble of physical systems that were prepared 
under the same complex of physical conditions, {\bf physical context.} However, by the conventional (orthodox Copenhagen)
interpretation of QM a wave function provides {\bf the complete description of an individual quantum system}.
Such a viewpoint, i.e., coupling of a wave function to an individual quantum system, induces hard problems 
in foundations of QM. One of them is the well known {\it problem of measurement}, see e.g. [3].

Another (less known, but not less important) problem is the problem of the origin of {\it quantum randomness.} 
By using the conventional interpretation of QM we could not apply the conventional ensemble approach to probability 
theory, see e.g. [4], [2]. Typically there are used such words as ``irreducible quantum randomness", see e.g. [4].
However, such irreducible randomness could not be explained by using conventional probability theory (Kolmogorov, 1933). 
Typically it is said that conventional probability theory does not work in quantum physics. We should use a new 
quantum calculus of probabilities based on linear algebra in the Hilbert space of quantum states. In fact such a
viewpoint is strongly supported by very strange (nonconventional) behaviour of quantum probabilities. 
It is well known that the classical rule for the addition of probabilities of  alternatives:
\begin{equation}
\label{eq:F1}
P=P_1+P_2
\end{equation}
does not work in experiments with elementary particles. Instead of this rule, we have to use quantum rule:
\begin{equation}
\label{eq:F2}
P=P_1+P_2+2\sqrt{P_1P_2}\cos\theta.
\end{equation}
The classical rule for the addition of probabilities of alternatives is perturbed by so called 
{\it interference term.}

Nevertheless, many famous physicists did not want to follow to the conventional (Copenhagen) 
interpretation of quantum mechanics. In particular, A. Einstein supported the ensemble viewpoint to a quantum state.
He wanted to preserve the conventional ensemble viewpoint even for quantum probabilities. The origin of his views can
be easily explained. In fact, in their pioneer papers M. Planck and A. Einstein used methods of classical statistical
mechanics (in particular, conventional probability theory) to describe {\it black body radiation} and
{\it photoelectric effect.} The only modification of classical statistical mechanics was due to the recognition 
of the principle (D) for 
the momentum variable. In some sense it was nothing 
than comeback to the purely corpuscular model of light (developed, for example, by Newton). 
However, a little bit later the purely corpuscular viewpoint to quantum systems became inconsistent
with statistical data obtained in {\it diffraction and interference experiments.} This statistical data
followed to quantum probabilistic rule (2) and not to conventional probabilistic rule (1).

To resolve this paradox between purely corpuscular behaviour of quantum systems in some experiments
and purely wave behaviour in other experiments, N. Bohr proposed the principle of {\it complementarity}, (C). By (C) there exists incompatible complexes of experimental physical conditions (contexts) determining incompatible physical observables, e.g. position and momentum. We remark that very often (C) is identified with Heisenberg's {\it uncertainty} principle. Of course, it is well known that N. Bohr proposed principle (C) after numerous discussions with W. Heisenberg on his uncertainty principle. However, (C) is essentially wider than Heisenberg's principle of mutual perturbations inducing incompatible observables.

We remark that fathers of the orthodox Copenhagen interpretation of QM from the very beginning were 
strongly against the possibility to reduce quantum randomness to classical randomness. From the beginning 
it was claimed that it is even in principle impossible to create a kind of deterministic (classical-like) model 
that would reproduce quantum probabilities via e.g. uncertainty in initial conditions and other parameters.
W. Heisenberg claimed this already in 1926 directly after the publication of his first paper on noncommutative 
representation of quantum of observables. Such a viewpoint was supported by N. Bohr in his discussions with
A. Einstein on the EPR paradox. It was the conventional viewpoint that quantum mechanics is {\it complete} and 
it would be impossible to provide finer description of physical reality.

As we know at present time, such a viewpoint was not justified. There exist classical-like models reproducing quantum probabilities. We can mention e.g. {\it Bohmian mechanics} and {\it stochastic electrodynamics.} Of course, these models are not free of problems. However, I do not think that any of these problems is essentially harder than e.g. the problem of measurement in the conventional Copenhagen approach. I think that immediate rejection of classical like prequantum models is merely a consequence of general prejudice against such models. For example, Bohmian mechanics is typically criticized for its 
{\it nonlocality.} However, at the same time it is widely accepted that conventional QM is nonlocal. 
Moreover, people totally forgot that Bell's arguments that nowadays are used to support quantum nonlocality
were originally proposed to support Bohmian mechanics. In fact, J. Bell was against {\bf locality} and 
not against {\bf realism}. Realism in the EPR-Bohm experiment (and, as a consequence, the possibility 
of deterministic description) was from the beginning evident for J. Bell
as the direct consequence of the precise anticorrelations. J. Bell wanted to prove nonlocality
of any deterministic prequantum model.\footnote{I would like to thank Shelly Goldstein who explained to me Bell's views to reality and locality.}

\section{CONTEXTUAL ENSEMBLE INTERPRETATION}
In the series of papers [5]-[7] I develop {\it contextual probabilistic approach} to QM. One of 
the main distinguishing features of this approach is {\it demystification} of quantum probabilities 
and, as a consequence, the whole QM. I started with (S) -- understanding of the fact that QM is just 
a special theory of statistical measurements. Then I tried to find a condition that would specify 
this theory of statistical measurements among all possible theories of statistical measurements. In particular, 
such a condition should distinguish QM and classical statistical mechanics (both considered as
theories of statistical measurements). Moreover, there should be a kind of 
correspondence principle that would establish transition from quantum statistics to classical statistics. 
The fundamental postulate of our model is the following one:

\medskip

{\bf ``Probabilities for all physical observables depend on complexes of experimental conditions under that 
measurements of observables are performed."}

We call this principle ``principle of contextual probabilities." This
principle of contextual probabilities is closely related to Bohr's principle of contextuality.
However, there is very important difference between those two principles. Bohr's principle
is a principle of {\it individual contextuality.} The basic notion related to Bohr's
principle of contextuality is the notion of individual {\it physical phenomenon.} Such a phenomenon is determined
by a complex of experimental physical conditions. My principle is a principle of {\it statistical 
contextuality.} The basic notion related to my principle is the notion of contextual probability.
Let us  consider the two slit experiment, 
see also further derivation of quantum probabilistic rule (2). For N. Bohr it is important 
that each individual spot on the registration screen 
depends on the whole experimental arrangement (e.g. both slits are open 
or only one of them). For me it is important that the probability distribution of spots
on the registration screen depends on the whole experimental arrangement. Of course, the reader
could argue that there is not so much difference between individual and statistical contextuality.
However, the situation is not so simple. Roughly speaking Bohr's individual contextuality does not 
imply that probability spaces used  in quantum theory should depend on complexes of experimental conditions.
Under Bohr's assumption on individual contextuality it is still possible (at least in principle)
to consider physical observables as random variables on one fixed probability spaces: various complexes
of physical conditions just determine various physical observables. In particular, after 75 years of Bohr's
individual contextualism it was still unclear why quantum probabilities follow to rules induced by the
Hilbert space structure and not by the structure of conventional probability theory.

In papers [5]-[7] it
was demonstrated (and it was surprising even for me) that starting with only
the principle of contextual probabilities we can derive quantum 
probabilistic rule (2). In its general form (2) is  quantum generalization of the well known 
formula of total probability. We discuss the contextual probabilistic interpretation of (2) on the basis
of the well known {\it two slit experiment.}

In the two slit experiment rule (2) is induced by
combining of statistical data obtained in three different experiments: 
both slits are open; only $j$th slit is open, $j=1,2.$ The main distinguishing feature
of statistical data obtained in these three experiments in the following one. 
By combining by (1) data
obtained in experiments in that only one of slits is open we do not get the probability distribution for
data obtained in the experiment in that both slits are open. On the other hand we never observe a particle
that passes through both slits simultaneously - it would be observed passing the first or second 
slit. There  is no the direct observation of particle splitting. As each particle passes  only one of slits,
we have the standard case of alternatives. Thus we should use conventional rule (1) for
the addition of probabilities of alternatives. This disagreement between experimental statistical
data and the rule of conventional probability theory looks as a kind of paradox. The traditional
solution of this paradox is the use of the wave model for elementary particles and, as a consequence,
the principle of complementarity. 
We now perform detailed contextual analysis for the two slits experiment. We consider the following
complexes of physical conditions, contexts.

${\cal S}=${both slits are open}, ${\cal S}_j=${only jth slit is open}, $j=1,2.$ 

In fact, probabilities in (2) are related to these three contexts.
Thus $P=P_{{\cal S}}(E)$ and $P_j= P_{{\cal S}\to {\cal S}_j} P_{{\cal S}_j}(E), j=1,2.$ 
Here we use various context-indexes. The $P_{{\cal S}}(E), P_{{\cal S}_j}(E)$ denote probabilities
of an event $E$ with respect to various contexts. The coefficients 
$P_{{\cal S}\to {\cal S}_j}, j=1,2,$ have another
meaning. In general these
are not probabilities of contexts ${\cal S}_j$ with respect to the context ${\cal S}$
(besides some very special, ``classical", situations), because the context
${\cal S}_j$ in general is not an event for the context ${\cal S}.$ In general we could 
not consider a complex of physical conditions ${\cal S}_j$ as an event in the probabilistic
space induced by some fixed context ${\cal S}.$

The coefficients $P_{{\cal S}\to {\cal S}_j}, j=1,2,$
are kinds of {\it splitting coefficients}. 
To be more precise, we consider some fixed source of particles and some fixed period $T$
during that we collect particles arriving to the registration screen. We consider an ensemble 
${\cal E}$ that consists of all particles that are collected on the registration screen during 
the period $T,$  when both slits are open. An  ensemble ${\cal E}_j$ consists of all particles that are collected on the registration screen during 
the period $T,$  when only $j$th slit is open. Suppose that ${\cal E}$ contains $N$ particles and
${\cal E}_j$ contains $N_j$ particles. Then splitting coefficients:
\begin{equation}
\label{eq:BP}
P_{{\cal S}\to {\cal S}_j}\approx \frac{N_j}{N},
\end{equation}
see [5] for the details. Here we use the symbol $\approx,$ since numbers $N$ and $N_j$ 
vary from one run (of duration $T)$ to other run (of the same duration). But if $N\to \infty,$
then fluctuations are negligibly small.

We remark (and it is important for our
further considerations) that we have the following statistical alternative condition:
\begin{equation}
\label{eq:BC}
P_{{\cal S}\to {\cal S}_1} + P_{{\cal S}\to {\cal S}_2} =1.
\end{equation}
The statistical alternative condition has the following meaning: the total number of particles that
arrive to the registration screen when both slits are open equals (in the average)
to the sum of corresponding numbers when only one of the slits is open. So by closing e.g. the first
slit we do not change the number of particles that pass the second slit (in the average).
In fact, (\ref{eq:BC}) gives the right description of alternative-situation in the two slit
experiment. It is not related to alternative passing of slits by a particle in the experiment
when both slits are open. This equation describes alternative sharing of particles between
two preparation procedures: $j$th slit is open, $j=1,2.$
However, the splitting coefficients $P_{{\cal S}\to {\cal S}_j}$  would not play 
so important role in our considerations. The crucial role will be played by contextual
probabilities $P_{{\cal S}}(E), P_{{\cal S}_j}(E).$

The conventional probability theory says
that, in fact,  we should  have:
\begin{equation}
\label{eq:FT}
P(E)= P({\cal S}_1) P(E/{\cal S}_1) + P({\cal S}_2) P(E/{\cal S}_2).
\end{equation}
There is assumed that complexes of physical conditions
${\cal S}_1$ and ${\cal S}_2$ could be considered as events with respect to probability space
containing $E$ as an event and 
$$
{\cal S}_1 \cap {\cal S}_2 = \emptyset  \; \mbox{and} \; {\cal S}_1\cup {\cal S}_2 ={\cal S}.
$$
This is the well known {\it formula of total probability.}
In many considerations (including works of fathers 
of quantum mechanics, see e.g. P. Dirac  and also R. Feynman) people 
set $P=P(E)$ and $P_j= P({\cal S}_1) P(E/{\cal S}_1).$
Finally, they get the contradiction between conventional
probabilistic rule (\ref{eq:F1}) (that in fact coincides with (\ref{eq:FT}))
and statistical data obtained
in the interference experiments and described by quantum rule (\ref{eq:F2}).

We would like to discuss physical and mathematical
assumptions used  in the conventional derivation of (\ref{eq:FT}).
The main physical assumption is that this formula is  derived for one fixed context ${\cal S}$ -
in the mathematical formalism - for {\it one fixed Kolmogorov probability
space}; see my book [2] on extended discussions on the role 
of the choice of Kolmogorov's probability space in quantum measurements.
To be precise, we have to write the conventional formula of total probability as 
\begin{equation}
\label{eq:FT1}
P_{{\cal S}}(E)= P_{{\cal S}}({\cal S}_1)  P_{{\cal S}}(E/{\cal S}_1)
+ P_{{\cal S}}({\cal S}_2)  P_{{\cal S}}(E/{\cal S}_2),
\end{equation}
taking into account that all probabilities are computed with respect to the same complex 
of physical conditions ${\cal S}.$

As we have already remarked, in Kolmogorov probability theory
it is also important that contexts ${\cal S}_1,  {\cal S}_2$ 
can be realized as elements of the field of events corresponding
to the context ${\cal S}.$
Thus we would get the contradiction between classical rule (\ref{eq:F1}) and
quantum rule (\ref{eq:F2}) only if we assume that splitting coefficients
$P_{{\cal S}\to {\cal S}_j}$ can be interpreted as 
$P_{{\cal S}}$-probabilities, $P_{{\cal S}\to {\cal S}_j}=P_{{\cal S}} ({\cal S}_j)$ 
and contextual probabilities $P_{{\cal S}_j}(E)$
as conditional probabilities with respect to the context ${\cal S}, P_{{\cal S}_j}(E)=
P_{{\cal S}}(E/{\cal S}_j).$  Here the conditional
probabilities are given by Bayes' formula
(the additional postulate of Kolmogorov's probability theory):
$P_{{\cal S}_j}(E)= P_{{\cal S}}(E \cap {\cal S}_j)/ P_{{\cal S}}({\cal S}_j) .$

But in general there are no reasons to assume that new complexes of conditions ${\cal S}_j$
are ``so nice" that new probability distributions are given by the Bayes' formula.
Thus, in general, the formula of total probability can be violated
when we combine statistical data obtained for a few distinct contexts.
In particular, it is not surprising that
this formula does not hold true in statistical experiments with elementary particles,
where the right hand side of this formula is perturbed by the interference term. 

\medskip

The following simple considerations give  the derivation of 
quantum probabilistic transformation (\ref{eq:F2}) in the 
classical probabilistic framework.  
Let ${\cal S}$ and ${\cal S}_j, j=1,2,$ be three
different complexes of physical conditions. We consider the transformation of probabilities induced by transitions 
from one complex of conditions to others: 
\begin{equation}
\label{eq:CT}
{\cal S} \rightarrow {\cal S}_1\; \mbox{and}\; {\cal S} \rightarrow {\cal S}_2.
\end{equation}
We start with introducing of splitting coefficients,
$P_{{\cal S}\to {\cal S}_j}.$  These are proportional coefficients for numbers of physical systems
obtained after preparations under the complexes of physical conditions ${\cal S}$ and ${\cal S}_j.$
If (starting with the same number of particles radiated by some fixed source)  we get $N$ and $N_j$ systems after  ${\cal S}$
and ${\cal S}_j$ preparations, respectively, then $P_{{\cal S}\to {\cal S}_j}$  are defined by
(\ref{eq:BP}). We assume that splitting coefficients satisfy to statistical alternative equation (\ref{eq:BC}).
This is quite natural conditions: splitting (\ref{eq:CT}) of the context ${\cal S}$ 
induces just  sharing of physical systems produced by a source. We have already
discussed this sharing in the two slit experiment. The same situation we have 
in neutron interferometry  for sharing of particles coming to detectors
when both paths are open and when just one of the paths is open.

We introduce  the {\it measure of statistical perturbations}
$\delta$ induced by context transitions:
$$
\delta({\cal S} \to {\cal S}_j;E)=P_{{\cal S}\to {\cal S}_1} [P_{{\cal S}}(E)-
P_{{\cal S}_1}(E)]+ P_{{\cal S}\to {\cal S}_2} [P_{{\cal S}}(E)-
P_{{\cal S}_2}(E)].
$$
This quantity describes the deformation of probability distribution $P_{{\cal S}}$
due to context transitions.
By using statistical alternative equation (\ref{eq:BC}) we get:
$P_{{\cal S}}(E)= P_{{\cal S}}(E) P_{{\cal S}\to {\cal S}_1}  + P_{{\cal S}}(E)P_{{\cal S}\to {\cal S}_2} .$

Thus we get:
\begin{equation}\label{eq:F3}
P_{{\cal S}}(E)= P_{{\cal S}\to {\cal S}_1}  P_{{\cal S}_1}(E)+ P_{{\cal S}\to {\cal S}_2}  P_{{\cal S}_2}(E)+
\delta({\cal S}\to {\cal S}_j;E).
\end{equation}
Transformation (\ref{eq:F3}) is the most general form of probabilistic transformations that can be induced
by context transitions. We can formulate a kind of  {\it{correspondence principle}} connecting  context unstable and  context
stable (``classical") transformations of probabilities: 

If ${\cal S}_j \rightarrow {\cal S}, j=1,2,$ 
i.e., $\delta({\cal S}\to {\cal S}_j;E) \rightarrow 0$, 
then contextual probabilistic transformation (\ref{eq:F3}) coincides (in the limit) with 
the conventional formula of total probability.

The perturbation term $\delta({\cal S}\to {\cal S}_j;E)$ depends on absolute magnitudes of probabilities. 
It would be natural to introduce normalized coefficient of the context transition 
$$
\lambda({\cal S}\to {\cal S}_j;E)=
\frac{\delta({\cal S}\to {\cal S}_j;E)}{2\sqrt{P_{{\cal S}_1/{\cal S}}  P_{{\cal S}_1}(E)P_{{\cal S}\to {\cal S}_2}  P_{{\cal S}_2}(E)}},
$$
that gives the relative measure of statistical deviations that can be induced by 
transitions from one complex of conditions, 
${\cal S},$ to others, ${\cal S}_j.$ Transformation (\ref{eq:F3}) can be written in the form:
\begin{equation}
\label{eq:F4}
P_{{\cal S}}(E)= \sum_{j=1,2} P_{{\cal S}\to {\cal S}_j}  P_{{\cal S}_j}(E)+
2 \; \sqrt{P_{{\cal S}\to {\cal S}_1}  P_{{\cal S}_1}(E)P_{{\cal S}\to {\cal S}_2}  P_{{\cal S}_2}(E)}\; \lambda({\cal S}\to {\cal S}_j;E)\;.
\end{equation}
In fact, there are two possibilities:

\medskip

1). $|\lambda({\cal S}\to {\cal S}_j;E)|\leq 1;$

2). $|\lambda({\cal S}\to {\cal S}_j;E)|\geq 1.$

In both cases it is convenient to introduce a new context transition parameter 
$\theta= \theta({\cal S}\to {\cal S}_j;E)$ and represent
the context transition coefficient in the form:
\[\lambda({\cal S}\to {\cal S}_j;E)=\cos \theta ({\cal S}\to {\cal S}_j;E) , \theta \in [0, \pi];\] 
and
\[\lambda({\cal S}\to {\cal S}_j;E)=\pm\cosh \theta ({\cal S}\to {\cal S}_j;E), \theta \in [0, \infty),\]
respectively.

We introduced a ``phase parameter" $\theta$ by purely mathematical 
reasons: to get a linear representation of probabilistic transformations 
induced by context transitions, see section 4. However, in some experiments it could 
occur that such a probabilistic parameter $\theta$ has some geometric meaning. This induces 
the illusion of a wave associated with a particle. However, in our contextual probabilistic
framework this is just a wave of probability. Such a ``wave" is associated not with an individual particle,
but with an ensemble of particles (in fact, with a transition from one ensemble to 
another).

We have two types of probabilistic transformations induced by the transition 
from one complex of conditions to another: 
\begin{equation}
\label{eq:F5}
P_{{\cal S}}(E)= \sum_{j=1,2} P_{{\cal S}\to {\cal S}_j}  P_{{\cal S}_j}(E)
+
2 \; \;\sqrt{P_{{\cal S}\to {\cal S}_1}  P_{{\cal S}_1}(E)P_{{\cal S}\to {\cal S}_2}  P_{{\cal S}_2}(E)} \cos \theta({\cal S}\to {\cal S}_j;E)\;.
\end{equation}
\begin{equation}
\label{eq:F6}
P_{{\cal S}}(E)= \sum_{j=1,2} P_{{\cal S}\to {\cal S}_j}  P_{{\cal S}_j}(E)
\pm
2 \;\;\sqrt{P_{{\cal S}\to {\cal S}_1}  P_{{\cal S}_1}(E)P_{{\cal S}\to {\cal S}_2}  P_{{\cal S}_2}(E)} 
\cosh \theta({\cal S}\to {\cal S}_j;E)\;.
\end{equation}
We derived quantum probabilistic rule (2)  in the classical probabilistic framework 
(in particular, without any reference to superposition of states) by taking into account context
dependence of probabilities.

Relatively large statistical deviations are described by transformation (\ref{eq:F6}). Such transformations do not 
appear in the conventional formalism of quantum mechanics. In principle, they could be described by so called 
{\it{hyperbolic quantum mechanics}}, [5].

\medskip

{\bf Conclusion.} For each fixed context (experimental arrangement), we have CLASSICAL STATISTICS.
CONTEXT TRANSITION induces interference perturbation
of the conventional rule for the addition of probabilistic alternatives.

\section{LINEAR ALGEBRA FOR PROBABILITIES}
One of the main distinguishing features of quantum theory
is the Hilbert space calculus for probabilistic amplitudes.
As we have already discussed, this calculus is typically associated
with wavelike (superposition) features of quantum particles.
We shall show that, in fact, the Hilbert space representation of
probabilities was merely a mathematical discovery. Of course, this discovery
simplifies essentially probabilistic  calculations. However, this is pure mathematics;
physics is related merely to the derivation of
quantum interference rule (2). 

The crucial point was 
the derivation (at the beginning purely experimental) of transformation (2)
connecting probabilities with respect to three different contexts. In fact, linear algebra
can be easily derived from this transformation. Everybody familiar with the elementary geometry
will see that (2) just the well known $\cos$-theorem. This is the rule to find
the third side in a triangle  if we know
lengths of two other sides and the angle $\theta$ between them:
$$
c^2 =a^2 + b^2 - 2 ab \cos \theta\;.
$$
or if we want to have "+" before $\cos$ we use so called {\it parallelogram law:}
\begin{equation}
\label{eq: P}
c^2 = a^2 + b^2 + 2 ab \cos \theta\;.
\end{equation}
Here $c$ is the diagonal of the parallelogram with sides $a$ and $b$ and the angle $\theta$
between these sides. Of course, the parallelogram law is  just the law of linear
(two dimensional Hilbert space) algebra: for finding the length $c$ of the sum ${\bf c}$
of vectors ${\bf a}$ and ${\bf b}$ having lengths $a$ and $b$ and the angle $\theta$ between them.

We also can introduce 
complex waves by using the following 
elementary formula: 
\begin{equation}
\label{eq:TTT}
a^2 + b^2 + 2 ab \cos\theta=|a+ b e^{i\theta}|^2\; .
\end{equation}
Thus the context transitions 
${\cal S}\rightarrow {\cal S}_j$ can be described by the wave of probability:
$$
\varphi=\sqrt{P_{{\cal S}\to {\cal S}_1}  P_{{\cal S}_1}(E)} 
+ \sqrt{P_{{\cal S}\to {\cal S}_2}  P_{{\cal S}_2}(E)}e^{i\theta({\cal S}\to {\cal S}_j;E)}.
$$

\section{ CONTEXTUAL PROBABILISTIC DERIVATION OF THE SUPERPOSITION PRINCIPLE IN THE TWO SLIT EXPERIMENT}
We shall study in more details the possibility of
contextual (purely classical) derivation of the superposition
principle for complex probability amplitudes, `waves', 
in the two slit experiment. We consider one dimensional
model. It could be obtained by considering the distribution of 
particles on one fixed straight line, very thin strip. It is supposed
that the source of particles is symmetric with respect to slits and 
the straight line (on the registration screen) pass through the center
of the screen. This geometry implies that coefficients
$P_{{\cal S}\to {\cal S}_j} =1/2, j=1,2.$ By the symbol $E_x, x \in {\bf R},$
is denoted the event of  the registration of a particle at the point $x$ 
of the straight line. We set:
$ p(x)= P_{{\cal S}}(E_x)$ and $p_j(x)= P_{{\cal S}_j}(E_x), j=1,2.$
These are probabilities that a particle  would arrive to the point 
$x$ (of the registration screen) under the complexes of physical conditions
${\cal S}$ and ${\cal S}_j,$ respectively. By using
(\ref{eq:F5}) we get:
$$
p(x)= \frac{1}{2} [p_1(x)+p_2(x)+ 2 \sqrt{p_1(x)p_2(x)} \cos \theta(x)] .
$$
This is just the special case of the general transformation of probabilities
induced by context-transitions.
By using (\ref{eq:TTT}) we represent this probability as the square of a complex
amplitude, $p(x)= \vert \phi(x)\vert^2,$ where 
\begin{equation}
\label{eq:S}
\phi(x)= \frac{1}{\sqrt{2}} (e^{i \theta_1(x)}\sqrt{p_1(x)} + 
e^{i \theta_2(x)}\sqrt{p_2(x)})
\end{equation}
and  probabilistic phases $\theta_j(x)$ are chosen in such a way that the phase shift
$\theta_1(x) - \theta_2(x) = \theta (x).$ We also introduce  complex amplitudes
for probabilities $p_j(x): \; \phi_j(x)= \frac{1}{\sqrt{2}} e^{i \theta_j(x)}\sqrt{p_j(x)}.$
Here $p_j(x)= \vert \phi_j(x)\vert^2.$ The complex amplitudes are 
said to be {\it wave functions:}  $\phi(x)$ is the wave function on (the straight
line of) the registration screen for both slits are open;
$\phi_j(x)$ is the wave function on (the straight
line of) the registration screen for $j$th slit  is open.

Let us set $\xi(x)=\frac{\theta(x)}{h},$ where $h>0$ is some scaling factor.
We have:
$$
\phi(x)= \frac{1}{\sqrt{2}} (e^{\frac{i \xi_1(x)}{h}}\sqrt{p_1(x)} + 
e^{\frac{i \xi_2(x)}{h}}\sqrt{p_2(x)})\; \mbox{and}\; 
\phi_j(x)= \frac{1}{\sqrt{2}} e^{\frac{i \xi_j(x)}{h}}\sqrt{p_j(x)}.
$$
By choosing $h$ as the Planck constant we get a quantum-like representation
of probabilities. We recall that we did not use any kind of wave arguments.
Superposition rule (\ref{eq:S}) was obtained in purely classical probabilistic
(but contextual!) framework.

Suppose now that $\xi$ depends linearly on $x: \xi_j(x)= \frac{{\bf p}_j x}{h}, \xi(x)= \frac{{\bf p} x}{h}, 
{\bf p} ={\bf p}_1 - {\bf p}_2.$ Under such an assumption we shall get interference
of two `free-waves' corresponding to momentums ${\bf p}_1$ and ${\bf p}_2.$ Of course, this
linearity could not be extracted from  our general probabilistic considerations. This is a
consequence of the concrete geometry of the experiment.

We underline that we did not have in mind to create some alternative to the standard calculus of
probabilities in Hilbert space. Our aim was demystification of this calculus. We demonstrated that 
we could use the conventional ensemble viewpoint to probability (as it was proposed by A. Einstein) 
even in experiments with quantum systems. However, we should not forget about the fundamental principle 
of contextual probabilities.

Finally, we remark that starting with this principle we obtain not only the standard 
trigonometric cos-like interference, but also hyperbolic cosh-like interference, see [5] for the details.
The latter calculus of (hyperbolic) probabilities could be realized (parallely to the standard quantum calculus) 
in the so called hyperbolic Hilbert space, see [5]. Another way to get a linear representation for the hyperbolic 
probabilistic transformation () is to use the formalism of Positive Operator Valued Measures, see e.g. [8]. Thus we 
can say that starting with very natural and clear postulate on contextual probabilities we can get the well known 
POVM-calculus, [9].

However, the reader can say: 

``Well, QM was demystified, by using contextual probabilities.
But, finally, you reproduced the standard probabilistic calculus in the Hilbert space 
(in its generalized POVM-form). Could you present some practical consequences of your 
contextual reduction to conventional probability theory?"

This question will be discussed in the next section.

\section{CONTEXTUAL VIEWPOINT TO QUANTUM INFORMATION}
What is the main difference between orthodox Copenhagen interpretation of QM and contextual 
interpretation (V\"axj\"o interpretation, [10])?

By the Copenhagen interpretation the origin of the very special quantum statistics is in the very
special behaviour of so called quantum systems. Here a wave function describes an individual system. Superposition is superposition of states for an individual system.

By the V\"axj\"o interpretation the origin of the very special quantum statistics is in the very
special complexes of physical conditions that are used to prepare ensembles of  physical systems.
Here a wave function (as it was claimed by Einstein) describes not an individual physical system, but an 
ensemble prepared under some complex of physical conditions. Superposition is not an individual property. 
It is related to statistical superposition in an ensemble. This is superposition of complexes of physical conditions.

The main fundamental consequence of the V\"axj\"o interpretation for quantum 
computing is that macroscopic classical physical systems could in principle exhibit 
quantum probabilistic behaviour under special complexes of physical conditions - special
preparation procedures for ensembles of such systems. In particular, the statistical superposition 
property could be induced by ensembles of macroscopic classical physical systems.
Thus so called {\it quantum parallelism} of computations could be 
realized by ensembles of macroscopic classical physical systems. 

Of course, the V\"axj\"o viewpoint to 
quantum parallelism differs crucially from the Copenhagen viewpoint. By the Copenhagen interpretation
after the act of quantum computation (unitary evolution) the quantum state of an individual physical system
contains all possible values of a Boolean function $f$ under computation. By the V\"axj\"o interpretation
all possible values of $f$ could be computed by performing computations over a large statistical ensemble of
physical system. By the V\"axj\"o interpretation the origin of the huge computational power of 
quantum computers is not in quantum parallelism,  
but it is a consequence of the possibility to prepare ensembles
of physical systems such that some fixed value of some parameter should be realized 
with probability that is close to probability 1. 
Therefore there is nothing mysterious that such a preparation could be in principle realized for macroscopic classical 
systems. 

Our main prediction for quantum computing is the possibility to create {\bf macroscopic classical quantum computer.}

{\bf Remark.} The combination ``classical quantum" looks as nonsense. 
However, we recall that here quantum is related only to special probabilistic behaviour of ensembles
physical systems.

At the moment we could not present any concrete model of such a classical/quantum computer.
However, we underline that quantum probabilistic behaviour could be found in various processes 
involving macroscopic systems, e.g. in economy.

Another consequence of the V\"axj\"o interpretation is the possibility to reduce quantum randomness to conventional 
ensemble randomness. This supports the original Einstein's viewpoint that QM is not complete and some finer description 
of physical reality is in principle possible.

Finally, we remark that the contextual probabilistic approach to Bell's
inequality [2], [11]-[13], demonstrated that there is no contradiction between the local realist description and violation 
of Bell's inequality by correlations calculated for special ensembles of physical systems (quantum systems). 
Contextual probabilistic analysis of Bell's argument induced new inequalities that are not violated by quantum
correlations, see [12], [13].

So if we follow to the V\"axj\"o interpretation of QM the problem of quantum nonlocality is still open! 
Of course, our arguments [2], [11]-[13],
could not be considered as arguments against nonlocality. It may be that QM is nonlocal! However, from the contextual probabilistic viewpoint Bell's arguments did no prove nonlocality of QM. There should be found some other arguments.

Our contextual viewpoint to quantum nonlocality has important consequences for quantum cryptography -
the only quantum information scheme that (at the moment) is well developed for real applications. 
The possibility to represent in principle quantum correlations by classical probabilistic integrals induces
some doubts in security of modern quantum information schemes.

Suppose that the classical reduction for quantum correlations would be
possible. Then the only special source of security of quantum cryptographic 
schemes would be high (classical) sensibility of quantum systems to external perturbations.

Acknowledgments:

I would like to thank L. Ballentine, S. Albeverio, E. Beltrametti,
T. Hida, D. Greenberger, S. Gudder, I. Volovich, W. De Muynck, J. Summhammer, P. Lahti, J-A. Larsson, H. Atmanspacher, 
B. Coecke, S. Aerts, A. Peres, A. Holevo,  E. Loubenets,  L. Polley, A. Zeilinger, C. Fuchs, R. Gill, L. Hardy,
B. Hiley, S. Goldshtein, A. Plotnitsky, A. Shimony, R. Jozsa, J. Bub, C. Caves, K. Gustafsson, H. Bernstein
for fruitful (and rather critical) discussions.

This investigation was  suppoted by EU-network "Quantum Probability with
Applications to Physics, Information Theory and Biology".

REFERENCES

 1. {\it Quantum Theory: Reconsideration
of Foundations,} A. Khrennikov, ed.,
Ser. Math. Modelling in Physics,
Engineering, and Cognitive Sc.,  V\"axj\"o University
Press, V\"axj\"o, 2002.\\
2. A. Yu. Khrennikov, {\it Interpretations of Probability,}
VSP Int. Sc. Publishers, Utrecht/Tokyo, 1999.\\
3.  P. Busch, M. Grabowski, P. Lahti, {\it Operational Quantum Physics,}
Springer Verlag, Berlin, 1995.\\
4.  A. Zeilinger, ``On the interpretation and philosophical foundations of
quantum mechanics",
in {\it  Vastakohtien todellisuus. Festschrift for K.V. Laurikainen,}
U. Ketvel et al., ed, pp. 53-60,  Helsinki University Press,  Helsinki, 1996.\\
5. A. Yu. Khrennikov, ``Linear representations of probabilistic transformations induced
by context
transitions", {\it J. Phys.A: Math. Gen.} {\bf 34}, pp. 9965-9981, 2001.\\
6. A. Khrennikov, ``Ensemble fluctuations and the origin of quantum probabilistic rule",
{\it J. Math. Phys.} 
{\bf 43}, pp. 789-802, 2002.\\
7. A. Yu. Khrennikov, ``Origin of quantum probabilities", 
{\it Quantum Probability and White Noise Analysis},
{\bf 13}, pp. 180-200, 2001.\\
8. A. Holevo, {\it Probabilistic and Statistical Aspects of Quantum Theory,}
Nauka, Moscow, 1980.\\
9. A. Yu. Khrennikov, E. Loubents, ``On relations between probabilities under quantum and classical
measurements", {\it Reports from MSI, V\"axj\"o University}, N. 02021, 2002.\\
10. A. Yu. Khrennikov, ``V\"axj\"o interpretation of quantum mechanics", preprint quant-ph/0202107.\\
11.  A. Yu. Khrennikov, ``Non-Kolmogorov probability models and modified Bell's
inequality", {\it J. of Math.} 
{\it Physics} {\bf 41}, pp. 1768-1777, 2000.\\
12.  A. Yu. Khrennikov, ``Statistical measure of ensemble nonreproducibility and correction to Bell's
inequality", {\it Il Nuovo Cimento} {\bf B 115}, pp.  179-184, 2000.\\
13. A. Yu. Khrennikov, ``A perturbation of CHSH inequality induced by fluctuations 
of ensemble
distributions," {\it J. of Math. Physics} {\bf 41}, pp.  5934-5944, 2000.
\end{document}